\documentclass[
aps,
amsmath,amssymb,
prl,
reprint,
groupedaddress,
superscriptaddress,
]{revtex4-2}

\usepackage{graphicx}
\usepackage{dcolumn}
\usepackage{bbm}
\usepackage{bm}
\usepackage{hyperref}
\hypersetup{colorlinks=true,citecolor=blue,urlcolor=blue,linkcolor=blue}
\usepackage[utf8]{inputenc}

\begin{document}


\title{Optical Fingerprint of Flat Substrate Surface and Marker-free Lateral Displacement Detection with Angstrom-level Precision}

\author{Shupei Lin}
\altaffiliation{These authors contributed equally to this work.}
\affiliation{School of Physics and Wuhan National Laboratory for Optoelectronics, Huazhong University of Science and Technology, Luoyu Road 1037, Wuhan, 430074, People's Republic of China}

\author{Yong He}
\altaffiliation{These authors contributed equally to this work.}
\affiliation{School of Physics and Wuhan National Laboratory for Optoelectronics, Huazhong University of Science and Technology, Luoyu Road 1037, Wuhan, 430074, People's Republic of China}

\author{Delong Feng}
\affiliation{School of Physics and Wuhan National Laboratory for Optoelectronics, Huazhong University of Science and Technology, Luoyu Road 1037, Wuhan, 430074, People's Republic of China}

\author{Marek Piliarik}
\affiliation{Institute of Photonics and Electronics of the Czech Academy of Sciences, Chabersk{\'a} 1014/57, 18251 Prague, Czech Republic}

\author{Xue-Wen Chen}
\email[To whom all correspondence should be addressed:\\ ]{xuewen\_chen@hust.edu.cn}
\affiliation{School of Physics and Wuhan National Laboratory for Optoelectronics, Huazhong University of Science and Technology, Luoyu Road 1037, Wuhan, 430074, People's Republic of China}
\affiliation{Institute for quantum science and engineering and Hubei key laboratory of gravitation and quantum physics, Huazhong University of Science and Technology, Luoyu Road 1037, Wuhan, 430074, People's Republic of China}

\begin{abstract}
We report that flat substrates such as glass coverslips with surface roughness well below 0.5 nm feature notable speckle patterns when observed with high-sensitivity interference microscopy. We uncover that these speckle patterns unambiguously originate from the subnanometer surface undulations, and develop an intuitive model to illustrate how subnanometer non-resonant dielectric features could generate pronounced {\color{black}interference contrast in the far field}. We introduce the concept of optical fingerprint for the deterministic speckle pattern associated with a particular substrate surface area and intentionally enhance the speckle amplitudes for potential applications. We demonstrate such optical fingerprints can be leveraged for reproducible position identification and  marker-free lateral displacement detection with an experimental precision of 0.22 nm. The reproducible position identification allows us to detect new nanoscopic features developed during laborious processes performed outside of the microscope. The demonstrated capability for ultrasensitive displacement detection may find applications in the semiconductor industry and super-resolution optical microscopy.
\end{abstract}

\maketitle

A substrate, such as a coverslip or wafer, is an indispensable component of any modern optical microscope. The interaction between the substrate and sample naturally imprints the substrate’s signatures into the observations. The substrate’s influence is especially pronounced in interference microscopes, including the phase-contrast microscope\cite{ZERNIKE1942974}, holographic microscope\cite{Gordon1966Science}, Nomarski microscope\cite{nomarski1950dispositif}, and interferometric reflection microscope\cite{Curtis1964JCB}. In particular, the interferometric scattering microscopy\cite{Vahid2004PRL,Taylor2019BookChapter,Young2019ARPC_Review,Kukura2009NatMethod}, as a recently established ultrasensitive imaging modality, directly relies on the interference between the reflected/transmitted field from the substrate surface and scattering of the sample. In principle, the inclusion of the substrate’s influence would not pose a problem as long as its topography and material are precisely known. Thus, typical microscope substrates used in the laboratory, such as glass coverslips, have been made extremely flat, often with surface roughness well below 0.5 nm\cite{Chada2015SciRep}. However, even employing such flat substrates, images obtained in interferometric scattering microscopy still exhibit seemingly mysterious  ``random'' speckle patterns, which have been speculated to originate from combined effects of uncontrollable imperfections including uneven illumination, glass inhomogeneity, surface roughness, dust, or stray light in the setup \cite{Taylor2019BookChapter,Kukura2016NatProt,Vahid2006OE,Hsieh2017ACSPhotonics}. The speckles set a lower limit for the size of detectable scatterers. While this limit can be partially bypassed in dynamically developing situations such as binding of proteins to the substrate where subsequent subtractions are applicable\cite{Vahid2014NC,Kukura2014NanoLett,Marek2019OpticsLaserTech,Marek2021SMTD}, the origin, influences, and implications of the speckle background have been poorly explored, and it remains challenging to detect stationary nanoscopic scatterers on the substrate despite their clear visibility under an atomic force microscope (AFM).

Here we unambiguously uncover that the dominant source of the speckles is the subnanometer undulations of the substrate surface. We develop an intuitive physical model to illustrate how subnanometer non-resonant dielectric surface undulations could generate pronounced {\color{black}interference contrast in the far field.} We introduce the concept of optical fingerprint for the deterministic speckle pattern associated with a particular substrate surface area. We demonstrate its applications for reproducible position identification and marker-free lateral displacement sensing with a precision of 0.22 nm. The position re-identification enables the detection of newly added nanoscopic features with ground-truth positions in the substrate. Lateral displacement sensing is crucial for a broad range of applications, including the localization of nano-fluorophores in super-resolution microscopies\cite{Stefan2007Science,Eric2006Science,Zhuang2006NatMethod}, and {\color{black}precision alignment of wafers during the multi-step nanofabrication processes} in the semiconductor industry\cite{Jinkook2014JMNN,Lee2011JMEMS}. 

\begin{figure}
	\centering
	\includegraphics[width=8.6cm]{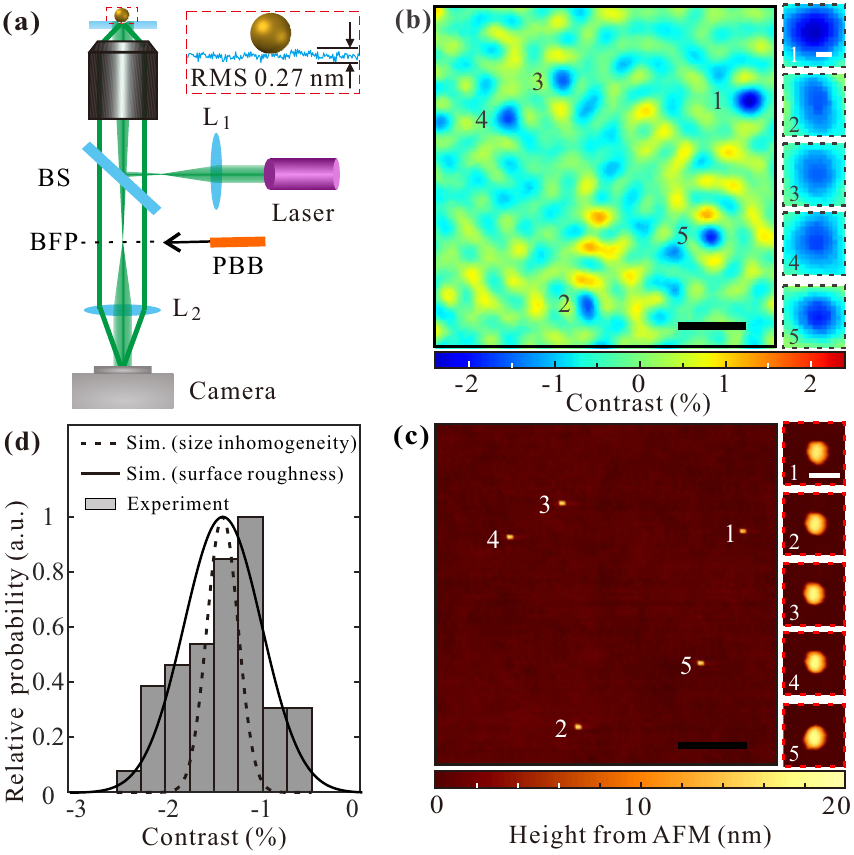}
	\caption{(a) Schematic of the experimental setup and the sample consisting of individual GNPs on a glass coverslip. A partial beam block (PBB) could be inserted to the back-focal plane (BFP). (b) Interference contrast image of the sample. Signals of the GNPs are labeled and enlarged on the right. (c) AFM topographic image of the same area of (b) with zoomed-in images of the GNPs on the right. (d) A histogram of measured interference contrasts from 52 GNPs on the same coverslip. {\color{black}The solid and dashed lines denote the relative probability distributions of the calculated contrasts with and without surface roughness considered (a.u. stands for arbitrary unit), respectively}. Scale bars: $1\ \mu m$ (black), 100 nm (white).
	}
	\label{FIG1}
\end{figure}

A schematic of a wide-field interferometric scattering microscopy setup is shown in FIG.\,\ref{FIG1}(a). A pulsed laser centered at 545 nm with a spectral width of 20 nm is launched into an oil-immersion objective (NA = 1.35) to illuminate the sample comprising a scattering specimen, such as single gold nanoparticles (GNPs), on a glass coverslip\cite{SM}. The scattering from the GNPs and the reflected illumination at the glass-air interface are collected by the microscope objective and imaged on the camera in a common-path interference arrangement. The interference contrast is defined as $c=I_{\rm t}/I_{\rm r} -1$, where {\color{black}$I_{\rm r}$ in principle is the reflection intensity from an ideally flat glass-air surface without the sample while $I_{\rm t}$  is the reflection intensity from the real interface with the sample. In practice $I_{\rm r}$ is approximately obtained with diminished influences from substrate surface roughness \cite{SM}.} FIG.\,\ref{FIG1}(b) displays the contrast image of the sample containing five isolated GNPs as indicated by the destructive interference dips. Magnified views of the dips are shown on the right. FIG.\,\ref{FIG1}(c) exhibits the AFM topographic image of the same sample area\cite{SM}. Contrary to the contrast image containing pronounced speckle patterns, the AFM image is extraordinarily clean with only the five GNPs observed in FIG.\,\ref{FIG1}(b). From zoomed-in images on the right of FIG.\,\ref{FIG1}(b) and \ref{FIG1}(c), for GNPs indexed from 1 to 5, the interference contrasts of -2.10\%, -1.40\%, -1.34\%, -1.47\%, and -1.60\%, and the corresponding heights of 16.4 nm, 17.0 nm, 17.2 nm, 17.8 nm and 18.1 nm, have been recorded. Since the vast majority of the GNPs are round in shape, the measured heights should represent the diameters. One thus observes that the measured contrast does not correlate to the GNP size according to the theory which predicts the contrast is proportional to the diameter in the third power\cite{Taylor2019BookChapter,Young2019ARPC_Review}. In particular, GNP \#1 has the smallest diameter but possesses the greatest contrast. A batch of 52 GNPs has been measured via both AFM and interferometric scattering microscopy\cite{SM} and a histogram of the interference contrasts is depicted in FIG.\,\ref{FIG1}(d). The dashed line denotes {\color{black}the relative probability distribution of the calculated contrasts for the 52 GNPs on a perfectly flat coverslip}\cite{SM,He2021JPD}. While the averaged contrasts between the calculations and measurements agree, the measured contrasts feature a much broader distribution. The phenomena of the abnormally broad contrast distribution in FIG.\,\ref{FIG1}(d), the speckle patterns in FIG.\,\ref{FIG1}(b) and the unexpected size-contrast correspondence, are intriguing and remain poorly understood.

\begin{figure}[t]
	\centering
	\includegraphics[width=8.6cm]{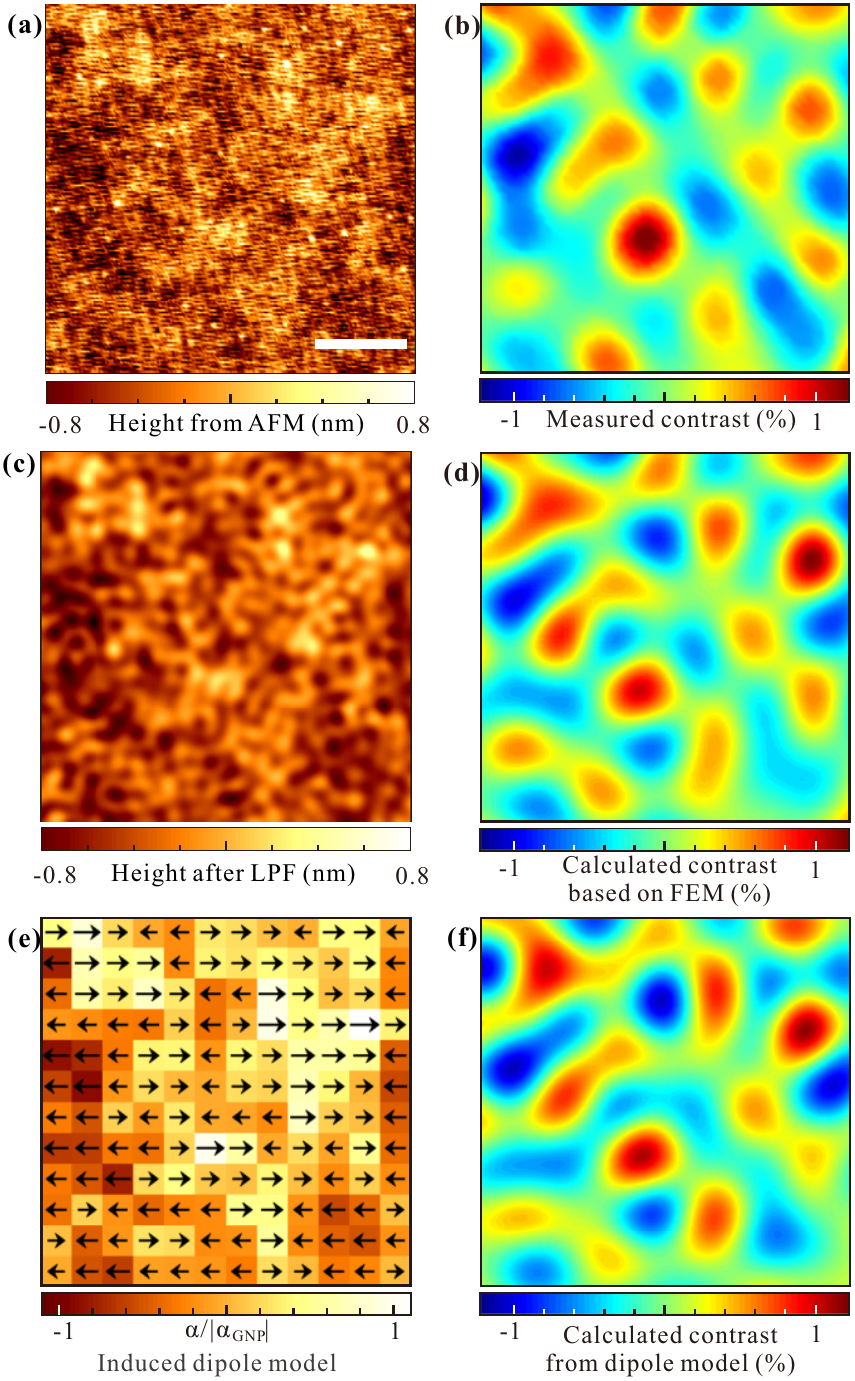}
	\caption{(a) AFM topographic image and (b) measured interference contrast image for the same coverslip area. (c) AFM image in (a) processed with a low-pass filter. (d) Simulated interference contrast based on (c). (e) Schematics of the induced dipole model and the distribution of the polarizability that is normalized by the absolute value of a 15 nm GNP. (f) Calculated interference contrast image based on the induced dipole model. Scale bar, 500 nm. 
	}
	\label{FIG2}
\end{figure}

To understand these phenomena, we scrutinize the situation of a bare glass coverslip. With delicate calibrations of the AFM measurements\cite{SM}, we manage to faithfully measure the coverslip surface topography with a height resolution better than 0.1 nm. FIG.\,\ref{FIG2}(a) and \ref{FIG2}(b) respectively display the measured topographic and interference contrast images for the same coverslip area. The coverslip is impressively flat with a roughness of only 0.27 nm (rms). Conversely, the optical image exhibits a contrast amplitude of about 1.2\%, which is comparable to that of a 15 nm-diameter GNP. It seems counterintuitive that a non-resonant dielectric surface with undulations of only 0.27 nm could generate such pronounced {\color{black}far-field interference contrast}. A closer examination of FIG.\,\ref{FIG2}(a) reveals the possible origin, \textit{i.e.}, the subnanometer undulations occur unevenly with some domains possessing more peaks than dips or vice versa, creating ``hill'' or ``valley'' domains laterally extending hundreds of nanometers. Considering such lateral scales, the hill or valley domains may effectively represent scattering volumes comparable to 10$\sim$20 nm-diameter nanoparticles, resulting in similar-level interference contrasts. To examine the hypothesis, the raw AFM image is processed with a low-pass filter (LPF) by removing spatial frequencies over $k_{\rm c} = 2\pi/(100\ \rm{nm})$, yielding the image of FIG.\,\ref{FIG2}(c). Indeed, hills and valleys emerge in terms of domains extending hundreds of nanometers laterally. The LPF processing facilitates rigorous numerical calculations of the scattering by surface roughness with an affordable computational demand. We employ the finite-element method (FEM) with a multiscale meshing scheme to resolve the subnanometer surface undulations in an area of $2\ \mu \rm{m} \times 2\ \mu \rm{m}$ \cite{SM}, and compute the interference image based on a full-wave electromagnetic theory\cite{He2021JPD,Forema2011JMO,Lalanne2016ACSPhotonics,Unlu2017BOE,Zhang2019Nanoscale}. FIG.\,\ref{FIG2}(d) presents the directly calculated interference contrast image based on the LPF-processed surface topography in FIG.\,\ref{FIG2}(c) under the plane-wave illumination. Imperfections such as the aberrations of various optical elements (including the microscope objective) have not been taken into account. The calculated contrast image agrees well with the measured one shown in FIG.\,\ref{FIG2}(b) not only in the speckle pattern but also in the contrast levels. These results unambiguously demonstrate that the coverslip surface with subnanometer roughness predominately generates the speckle contrast pattern.

The calculated contrast pattern is robust against the choice of $k_{\rm c}$ if $k_{\rm c} \textgreater NA \times k_0$, with $k_0$ being the vacuum wavenumber\cite{SM}. This is understandable because $NA \times k_0$ is the lateral wavenumber achievable by our optical system. This observation leads us to draw an induced dipole model. As depicted in FIG.\,\ref{FIG2}(e), we divide the surface area in FIG.\,\ref{FIG2}(a) into $12 \times 12$ grids and approximate each grid as an electric dipole induced by the illumination. Then we consider each grid as a dipolar scatterer with a polarizability $\alpha _i=(\epsilon  _g-\epsilon  _0)V_i$, where $\epsilon_g$ and $\epsilon_0$ are the dielectric permittivities of glass and vacuum, respective. $V_i$ is the volume of the $i$-th domain as $V_i=\iint_{S_i} h dS$ with $h$ being the measured height,which encodes the information of local surface topography into the dipole. Apparently, the ``hills'' give positive $V_i$ while ``valleys'' possess negative $V_i$. FIG.\,\ref{FIG2}(e) displays the distribution of the dipoles’ polarizabilities normalized by the absolute polarizability of a 15 nm-diameter GNP, which varies from -1.1 to 1.1, consistent to our estimations. The contrast image calculated from the induced dipole model\cite{SM} is displayed in FIG.\,\ref{FIG2}(f) and closely matches the pattern shown in FIG.\,\ref{FIG2}(d). This picture enables us to explain the experimental results in FIG.\,\ref{FIG1}(b) and \ref{FIG1}(d). The GNPs reside above different locations of the coverslip, \textit{i.e.}, induced dipoles of various polarizabilities (FIG.\,\ref{FIG2}(e)). The superposition of the induced dipoles from the GNP and the rough-surface domain beneath gives the total dipole that varies with location. This explains the previously observed unexpected GNP size-contrast correspondence. We have simulated a 17.3 nm GNP (average size from measurement) at various locations of the substrate through FEM simulations\cite{SM} and presented {\color{black}the relative probability distribution} of the calculated contrasts in a solid line in FIG.\,\ref{FIG1}(d), which nicely agrees with the measured histogram.

\begin{figure}[t]
	\centering
	\includegraphics[width=8.6cm]{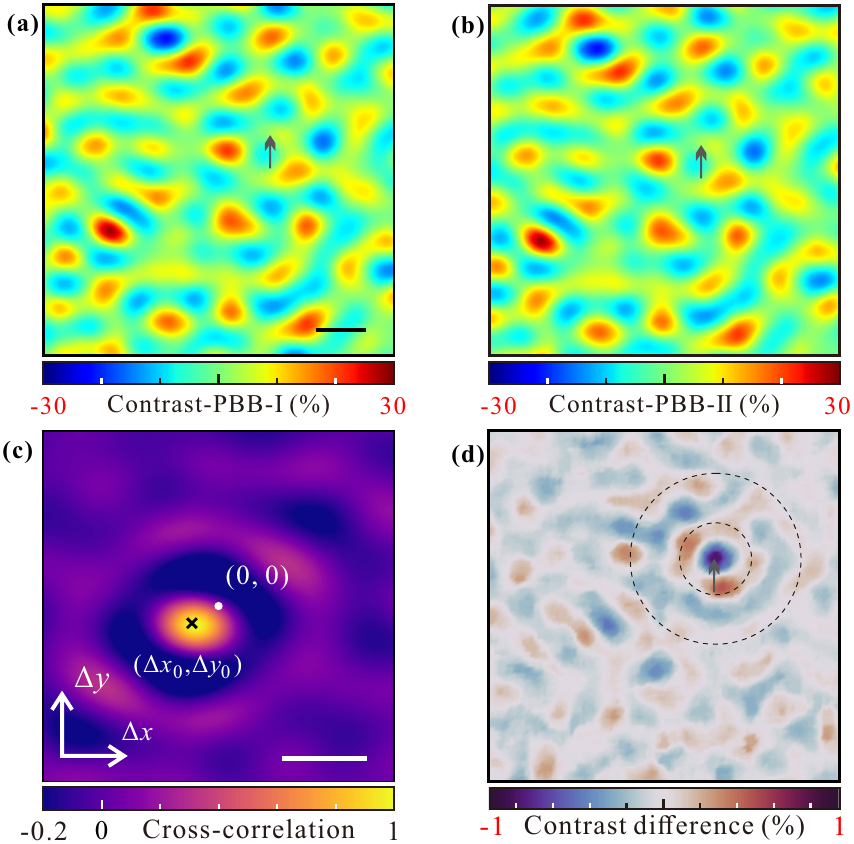}
	\caption{Interference contrast images for the same coverslip area (a) before and (b) after spin-coating 5 nm GNPs. A PBB is placed at the BFP to amplify the contrast. (c) Normalized cross correlation between image (a) and image (b) as a function of lateral displacement $(\Delta x, \Delta y)$. The white dot and black cross denote zero displacement and a displacement of $(\Delta x_0,\Delta y_0)=(-166.8 \ \rm{nm}, -90.4 \ \rm{nm})$ giving the maximum correlation. (d) Contrast image obtained from subtracting (a) from (b) displaced by $(\Delta x_0, \Delta y_0)$. The dashed circles indicate the point spread function of a 5 nm GNP. Scale bars, 500 nm.
	}
	\label{FIG3}
\end{figure}

Our findings illustrate that an ultraflat substrate surface area with roughness well below 0.5 nm does not possess translational invariance and instead features a deterministic speckle pattern serving as its optical fingerprint. In the following, we show that the optical fingerprints can be enhanced and leveraged for lateral position and displacement sensing with subnanometer precisions. Under wide-field illumination, the scattering and reference beams are spatially separated at the back focal plane (BFP) and thus the reference can be selectively attenuated by a partial beam block (PBB) (FIG.\,\ref{FIG1}(a)) to enhance the interference contrast\cite{Kukura2017ACSPhotonics,Liebel2017NanoLett,Unlu2017Optica,Hsieh2019Nanoscale,Marek2021NC}. In our experiment, the PBB is a 1 mm-diameter thin metal film (10 nm Ti $\backslash$110 nm Au) on a glass plate and amplifies the interference contrast by nearly 20 times\cite{SM}. 

To demonstrate reproducible lateral position identification, we perform microscopy measurements on a new glass coverslip and record its contrast pattern (Contrast-PBB-I) as in FIG.\,\ref{FIG3}(a). Next, we remove the coverslip from the setup, immobilize individual 5 nm GNPs to it via spin-casting the colloidal solution, and then remount it back to the microscope. While the axial focus can be readily recovered, laterally the coverslip can only be coarsely adjusted to the original position by inspecting the wide-field image. The measured interference contrast image (Contrast-PBB-II) is displayed in FIG.\,\ref{FIG3}(b). There are two features apparent from the comparison of FIG.\,\ref{FIG3}(a) and \ref{FIG3}(b), \textit{i.e.}, (i) the two patterns are nearly identical except for a lateral displacement; (ii) there is no indication in Contrast-PBB-II suggesting any GNP immobilized on the substrate. We evaluate the cross-correlation between Contrast-PBB-I and -II as a function of lateral shift $(\Delta x, \Delta y)$ according to $\sigma_{\rm{XC}}(\Delta x,\Delta y)=\left \langle c_{\rm{\uppercase\expandafter{\romannumeral1}}}(x,y) c_{\rm{\uppercase\expandafter{\romannumeral2}}} (x+\Delta x,y+\Delta y) \right \rangle/\sqrt{\left \langle c_{\rm{\uppercase\expandafter{\romannumeral1}}}^2 (x,y) \right \rangle \left \langle c_{\rm{\uppercase\expandafter{\romannumeral2}}}^2 (x,y) \right \rangle}$, where $\left \langle \cdots \right \rangle$ represents the integration about $(x, y)$ over the area while $c_{\rm{\uppercase\expandafter{\romannumeral1}}}$ and $c_{\rm{\uppercase\expandafter{\romannumeral2}}}$ denote Contrast-PBB-I and Contrast-PBB-II, respectively. The calculated correlation shown in FIG.\,\ref{FIG3}(c) features a symmetric elliptical peak. By fitting the peak with a two-dimensional Gaussian function, we determine its center at $(\Delta x_0,\Delta y_0)=(-166.8 \ \rm{nm}, -90.4 \ \rm{nm})$ with standard deviations of (0.53 nm, 0.43 nm) \cite{SM,Thompson2002BioJour}, which is the lateral displacement of the coverslip in FIG.\,\ref{FIG3}(b) with respect to FIG.\,\ref{FIG3}(a). With the known displacement, we align the sample back to its original position and obtain the difference between aligned Contrast-PBB-II and Contrast-PBB-I shown in FIG.\,\ref{FIG3}(d). The resulting pattern shows two symmetric interference rings highlighted by the dashed circles and allows the localization of newly immobilized nanoscopic feature with the ground-truth position of the substrate indicated by the speckles. We identify a 5 nm GNP on the substrate, whose signal is about 30 times smaller than the speckle background. The unprecedented {\color{black}observation is further supported by additional correlated optical and AFM topographic measurements \cite{SM}.} We note that previous reports of detecting similar or smaller particles were performed in a solution environment and in dynamically developing situations where sequential subtractions of acquired images can be applied within short temporal delays to reveal changes due to each particle binding.

\begin{figure}
	\centering
	\includegraphics[width=8.6cm]{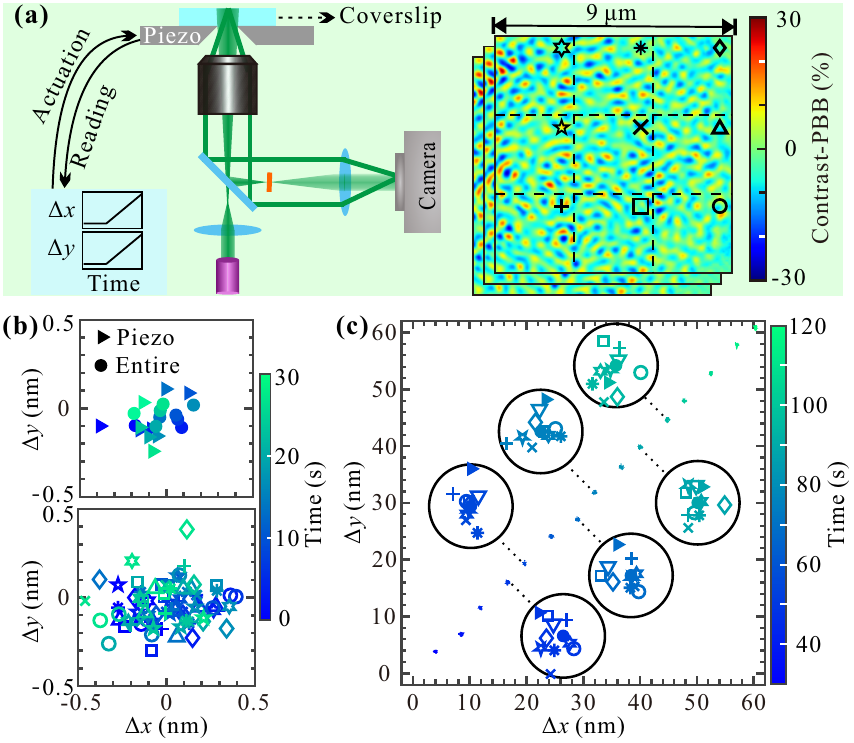}
	\caption{(a) Illustration of the procedure for lateral displacement detection and benchmark (see text). Symbols in the contrast image denote the various regions. {\color{black}(b) Displacements from readings of nanopositioner and from optical fingerprints of the entire area and various regions when the nanopositioner is kept still.} (c) Measured displacements when the nanopositioner is actuated to stepwisely (4 nm) move in each direction. Symbols in the solid circles are the zoomed-in views of the data points and the circles have the radii of 0.5 nm.
	}
	\label{FIG4}
\end{figure}

Next, we demonstrate the fingerprints of the substrate allow for ultrasensitive lateral displacement detection. As sketched in FIG.\,\ref{FIG4}(a), we image a $9\ \mu \rm{m} \times 9\ \mu \rm{m}$ area of a glass coverslip when it is stepwise displaced using a closed-loop piezo nanopositioner (P-621.2CL, Physik Instrumente). {\color{black}We divide the imaged area into 9 regions (see the symbols) of the same size, which show very different and mutually uncorrelated speckle patterns (see FIG.\,\ref{FIG4}(a)). For every step, speckle patterns of each region at the two time-stamps are cross-correlated to evaluate the displacement. The cross-correlation profiles are unique for each region \cite{SM}. Thus they constitute a collection of 9 independent and simultaneous realizations of optical displacement detection for the same move.} To benchmark the optically detected displacements, we collect the real-time readings of the capacitance probe of the nanopositioner\cite{SM}. As shown in FIG.\,\ref{FIG4}(b), while the nanopositioner remains stationary for 30 seconds, the readings from the nanopositioner (averaged in 3 seconds) and the optically detected displacements both reside around zero displacements with standard deviations of 0.18 nm and 0.27 nm, respectively. The results indicate good stability of our home-built microscope as well as the performance of the piezo stage position readout. Therefore we argue that, any other uncontrolled displacement of the sample is negligible at the time scales of the experiment and any move of the coverslip will be due to the nanopositioner. FIG.\,\ref{FIG4}(c) summarizes the measurements with color-coded symbols for the situation when the nanopositioner is actuated to move in a stepwise manner in both $x$ and $y$ directions (step size: 4 nm in each direction). The color scale indicates the elapsed time. Six of the data points are zoomed in as the insets, where the solid circles have the radii of 0.5 nm. All the averaged readings of the nanopositioner exhibit excellent agreement with the optically detected displacements. Crucially,independent optical displacement evaluations denoted by various symbols give the same moving trajectory and the standard deviation of the independent realizations for the 15 measured steps is 0.22 nm. 

{\color{black}Lateral displacement sensing promises a range of important applications including precision alignment for wafer nanofabrication in the semiconductor industry. Existing approaches rely on optically detectable markers such as nanoantennas or inscribed nanostructures with a possible combination of vector beam illumination \cite{Urbach2015PRL,Banzer2016NC,Banzer2018PRL,Wei2018PRL,Tischler2018ACSPhotonics,Yuan2019Science,Xi2020PRL,Banzer2020NC,Xi2020PRApplied,Zhan2009AOP}. Our scheme is marker-free, relies on naturally existing substrate roughness and operates under a simple wide-field illumination. Light sources such as a diode laser could also be used for illumination\cite{SM}. Thus, our technique should be directly applicable for fine alignment for various types of substrates. In particular, it potentially allows for compensation of temperature-induced drift at any designated position of a wafer-sized substrate\cite{Lee2011JMEMS} without the need for any adjacent alignment markers.}

We have uncovered that subnanometer substrate surface undulations are the dominant source of the speckle background observed in high-sensitivity interference microscopy. We show the unevenly-distributed subnanometer roughness could give rise to an induced dipole comparable to a 15 nm GNP. The induced dipoles with surface-topography-encoded polarizabilities form the optical fingerprint of the substrate surface, which can be repeatedly identified and localized in experiments. Moreover, the speckle amplitudes can be enhanced to about 30\% of the reference field. These experiments vividly show how subnanometer non-resonant dielectric features could generate prominent far-field optical responses. We've demonstrate their potential applications for transverse position and displacement sensing with angstrom-level precisions. The precise positioning of the substrate and the complete recovery of the measured speckle pattern allowed us to identify and localize nanoscopic features (with contrast $\sim 1/30$ of the speckle background) immobilized on the surface during processes performed outside of the microscope, which offers great flexibility and new possibilities for the use of substrate in ultrasensitive detection. The capabilities of marker-free precise alignment and localization of immobilized nanometer features demonstrated here are essential for a myriad of modern technologies spanning from super-resolution imaging to advanced nanofabrication in the semiconductor industry.

We acknowledge financial support from the National Natural Science Foundation of China (Grant Number 12011530395, 11874166), Science and Technology Department of Hubei Province,China (Project 2022BAA018), Czech Grant Agency (Project 22-11753S) and the Czech Academy of Sciences under the bilateral collaboration project NSFC-21-18. We also thank J. Tang for useful discussions and N. Jiao for assistance with the experiment.




%

\end{document}